# Levitation? Yes, it is possible!
## (Additional material)*

Alberto T. Pérez[1], Pablo García-Sánchez, Miguel A. S. Quintanilla and Armando Fernández-Prieto

Departamento de Electrónica y Electromagnetismo, Universidad de Sevilla, Avda. Reina Mercedes, s/n, 41012 Sevilla, Spain.

**\*This document contains supplementary material for the article with the same title that appeared in American Journal of Physics in 2019.**

I. INTRODUCTION

High schools students from Spain attended a scientific summer camp at the University of Seville in July 2018. The students worked in a project with the main objective of learning the basic laws of magnetism while pursuing a challenging goal: to achieve stable magnetic levitation.

The students participate in eight sessions. The first two are a quick theoretical review, the following five sessions are devoted to experimental work at the laboratory. The last session of the scientific camp consists in an oral presentation of the results. Table I shows the project schedule. For the sake of brevity, we briefly described the experimental sessions in the main article published in American Journal of Physics [1]. This supplementary material contains further information of the experimental work and results obtained by the students.

TABLE I: WEEK SCHEDULE

|           | 9.30-11.30 h                      | 12.00-14.00 h          | 15.30-17.30 h |
|-----------|-----------------------------------|------------------------|---------------|
| Monday    | Welcome                           | Theory and demostrations |              |
| Tuesday   | Force between magnets             | Diamagnetic levitation |               |
| Wednesday | Jumping ring and Alcon levitator  | Superconductors        | Levitron      |
| Thursday  | Presentation elaboration          |                        |               |
| Friday    | Rehearsal                         | Presentation           |               |

[1] Corresponding author: alberto@us.es

## II. EXPERIMENTAL SESSIONS

### A. *Limitation of degrees of freedom*

The first experimental way of achieving levitation that we propose to the students is the restriction of the number of degrees of freedom of the magnets. During the first experimental session, the students measure the repulsive force between two hollow disk magnets with like poles facing. The magnets are inserted in a plastic rod as part of a demonstration set (from "Antigravity magnetic levitation" a toy by Kidzlabs). The top magnet is loaded with a plate containing glass beads until the distance between the magnets reaches a certain value. In this way, the students measure the force as a function of the distance between the two magnets. The beads weigh is measured with a digital balance (See Fig.1 in the main article). Fig. 1 is a typical plot obtained in this way.

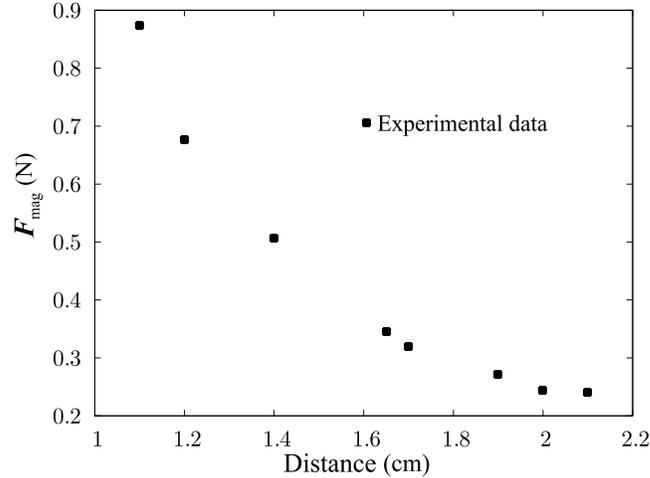

Fig. 1. Magnetic force as a function of distance between magnets.

The plot in fig.1 shows that the magnetic force decreases with distance. This can be used to explain the concept of stable equilibrium: If the floating magnet approaches the bottom magnet, the repulsive force becomes greater than the weight and pushes the magnet upwards. On the contrary, if the floating magnet is lifted, the repulsive force becomes smaller than its weight and tends to fall. It is then found that the total force on the floating magnet acts as a restoring force. Consistently, oscillations of the magnet position occur when displaced from its equilibrium point.

During this session, we allow the students to play with the set of magnets. The aim is that the students experience that the equilibrium is only possible by restricting the motion of the magnets. As soon as the sustaining pole is removed, the magnets flip and become stuck by the opposite poles. In order to introduce the students to some instrumental techniques, this experimental session is completed with the measurements of the magnetic field of a bar magnet as a function of the distance to the magnet pole. The measurements are made with a Hall probe and a teslameter, supplied by Phywe. A typical curve is shown in Fig. 2. As expected, the magnetic field decays as $1/r^2$ near the pole, and as $1/r^3$ far from it ($r$, being the distance between the pole and the probe).

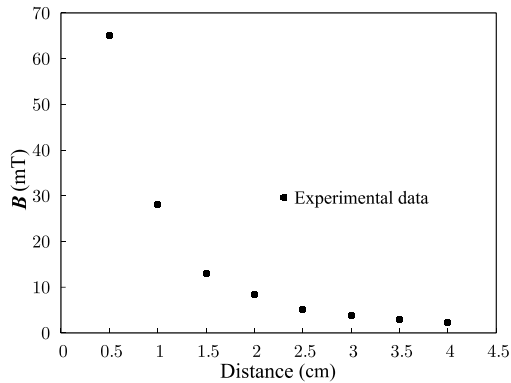

Fig. 2. Magnetic field as a function of distance of the bar magnet to the magnetic pole.

*B. Diamagnetic levitation*

The second experimental session is focused on diamagnetic levitation. To this end, we provide the students with the following material (see Fig. 2 in the main article):

- A set of disk shape neodymium magnets
- A metallic (iron) piece with angular shape making a 90 degrees angle and 120 mm long. This piece is intended to aid the magnets positioning
- Two supports made of aluminum
- A small compass-like device. This compass is used to determine magnet poles
- A graphite mine 60 mm long

The students are encouraged to find a configuration in which the graphite mine levitates freely [2]. All students groups succeeded in this experience and, in general, found several arrangements of the magnets resulting in a stable mine levitation. Given the small gap between the graphite piece and the magnets, students use a binocular microscope for inspection of the levitating bar (see Fig. 3 in the main article).

*C. Superconductors*

Superconductors behave as superdiamagnetic materials, that is to say, when subjected to a magnetic field they generate electric currents that cancel out the field inside the material. Thus, superconductors are said to behave as magnetic mirrors, i.e. the magnetic field outside the material can be described as originated by a magnet identical to the external one but with like poles facing, see fig. 3. This "image" magnet continuously follows the external magnet – a phenomenon that can be used to levitate the latter [3, 4].

In this experimental session we use a disk shape piece of type II superconductor (YBaCuO) with a critical temperature of 95 K, above the nitrogen liquefaction temperature (77 K). We use liquid nitrogen to cool down the superconductor disk below its critical temperature. For safety reasons, the students are not allowed to handle liquid nitrogen and the experiment is carried out by one of us. We place the YBaCuO disk on an expanded polystyrene container and poor liquid nitrogen until the superconductor is completely covered. After a couple of minutes to allow the superconductor to cool down, we place a small magnet above the superconductor disk and levitation occurs as a consequence of the Meissner effect.

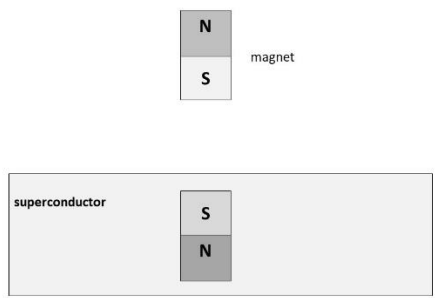

Fig. 3. A superconductor behaves like a magnetic mirror.

## D. Induced current: Jumping ring

Levitation by induced current is demonstrated with two experimental setups. The first demonstration is the jumping ring, also known as Thomson's ring [5, 6, 7]. The experimental setup consists in a solenoid with 1200 turns, a ferromagnetic core (we use three pieces, each 10 cm long, stacked), a multimeter in the ammeter function, a switch, a set of cables, a ruler and a set of aluminum rings with different heights (Fig. 4).

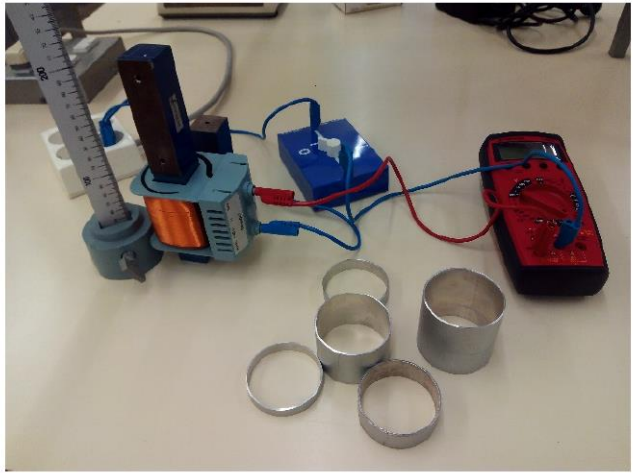

Fig. 4. Experimental setup for jumping ring experiment.

The solenoid is connected to the AC mains through the multimeter and the switch. One of the ring is placed over the solenoid and concentric with the iron core. After closing the circuit with the switch, a sudden current pulse is induced on the ring and the magnetic force on that current throws the ring up to a certain distance. Another possible experience is to switch first the current on and then place the ring on the solenoid. Continuous levitation of the ring is achieved since the iron core prevents it from lateral displacements.

Despite the simplicity of this experimental setup, its interpretation requires of several concepts, some of them too complex for high-school students. The sudden jump can be interpreted, qualitatively, in terms of induced current and Lenz's law: the induced current opposes the increasing magnetic field in the ring, therefore the ring reacts as a magnet with its poles reversed to those of the solenoid. As a consequence, it appears a transit repulsive force that produces the ring jump. However, in the steady state the current on the solenoid is an AC one. Thinking on terms of induced current and Lenz's law, the magnet poles are reversed, when the current increases, but have the same orientation, when the current decreases. Therefore, the force would be repulsive during half cycle and attractive during the other half. The ring levitates during the steady sinusoidal due to the lag phase between the solenoid current and the induced current due to the ring self-induction [5]. This reasoning is too far-reaching for our high-school teenagers and we decided to keep things at the empirical level. We just ask them to measure the levitation height as a function of the solenoid current and the ring height. Typical results are shown in Fig. 5.

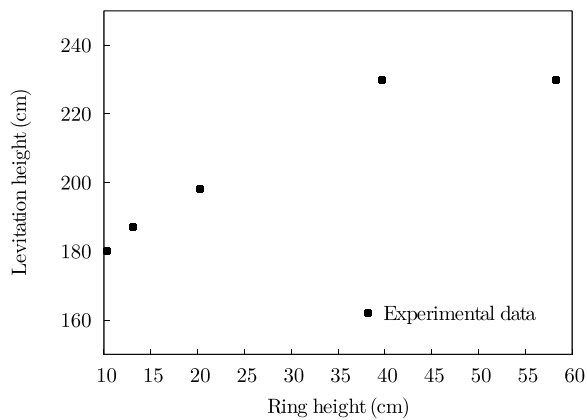

Fig. 5. Levitation height as a function of ring height when a 1.1 An AC current is applied to the solenoid in the jumping ring experiment.

### E. Induced current: the Alcon levitator

The second demonstration of levitation by induced current is Alcon's levitator [3, 8]. Fig. 4 in the main article shows a scheme of the geometric configuration: two metal cylinders rotate at the same angular speed but opposing directions, and a magnet is placed above cylinders. The induced current on the cylinders exerts a vertical force on the magnets pointing upwards. This magnet can be levitated if the rotation of the cylinders is fast enough.

Very high rotation speeds are required for the levitation of the magnet. For safety reasons, we do not propose that our students attempt a levitation experiment. Instead, with the aim of quantifying the force the on the magnet, they measure the apparent magnet weight for small angular speed of the cylinder. From these measurements the students extrapolate the minimum speed required for levitation.

Fig. 6 shows the experimental setup. A voltage supply is used to control the voltage of two DC motors that drive the cylinders rotation. A photovoltaic shutter connected to an oscilloscope is employed to measure the cylinder angular speed. A neodymium magnet is attached to a string and hanging from a laboratory sliding weight scale. By using a sliding weight scale, the magnet position with respect to the cylinders remains the same after the weight scale is balanced. The magnet is maintained within the walls of a transparent plastic tube for avoiding large horizontal displacements.

During this session the students proceed as follows. The voltage supply is set to a given voltage, ranging from 10 to 35 volts, the velocity of the cylinders is measured in the oscilloscope. Then, the apparent weight of the magnet is measured in the scale. The measurement is repeated for voltages within the available range. Fig. 7 shows a typical plot of the measurements. It can be extrapolated that the apparent weight vanishes for a velocity around 4.300 rpm and, thus, levitation could occur. After the students complete the measurements and analysis of the results, we perform a levitation experiment with more powerful motors connected to the cylinders and taking appropriate safety measures.

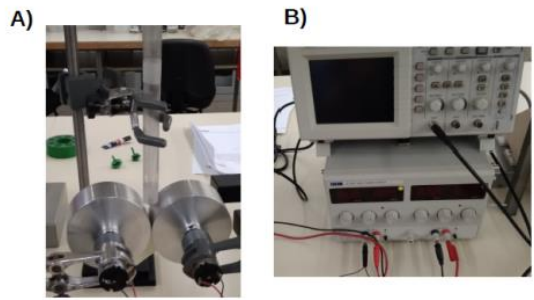

Fig. 6: Experimental setup for the Alcon levitator measurements.

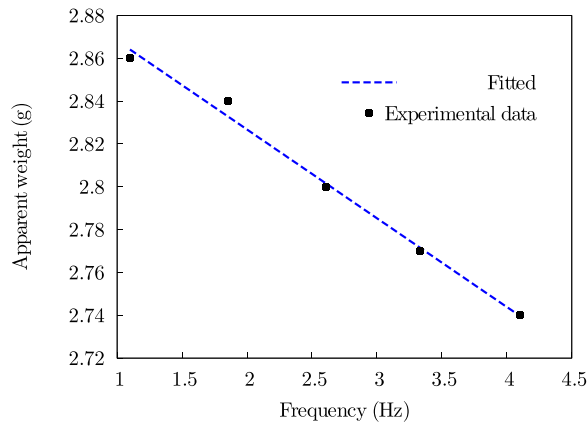

Fig. 7. Apparent magnet weight as a function of cylinder angular velocity for the Alcon levitator.

*F. Levitron*

The last experimental session is dedicated to Levitron, a toy invented by Roy Harrigan and commercialized by Fascinations. This toy consists of a magnetic top that hovers above a magnetic base. The top levitation arises from the magnetic repulsion with the base. The gyroscopic effect prevents the flipping of the top, although stable levitation is only achievable within a certain range of angular velocities. The top mass must be precisely adjusted since a top too light flies away, while a heavy one falls down. The Levitron toy is accompanied with a set of washers of different weights that aid to adjust the top mass.

The stability of Levitron is a complicated mathematical problem that has been addressed by several authors [10,11,12,13]. For this experimental session, our goal is that the students understand the stability in the vertical direction and realize the importance of the gyroscopic effect. Previous to the work with the Levitron, the students experience the stabilizing effect of a rotary motion by playing with a giant spinner made out of a bike wheel with handles on both axle's ends. The existence of a mechanism of horizontal stabilization can be shown by playing with a magnetic top with a magnetic moment weaker than the one supplied with the Levitron. The body of this weakly magnetized top was manufactured by 3D printing, with a weak ring magnet glued to it. This customized top was easier to spin on the Levitron than the original one and, although incapable of levitating, it drifted readily to the axis of revolution of the Levitron's base, whereas when it was spun on a table top it drifted randomly of the table's surface.

After these preliminary activities, the students are encouraged to try to levitate the top. Only 3 out of 31 students that attended the camp succeeded to levitate the top for longer than 15 s. While the students understood that weight and magnetic force must balance each other, few realized how sensitive the Levitron stability is to the top weight and changed the top weight almost at random, rather than following a systematic search by increasing the weight in small increments. Besides most students failed to lift the base slowly enough to smoothly place the top at the equilibrium height.